\documentclass[a4paper, USenglish, cleveref, autoref, thm-restate]{lipics-v2021}

\usepackage{listings}
\DeclareFontShape{OT1}{cmtt}{bx}{n}{<5><6><7><8><9><10><10.95><12><14.4><17.28><20.74><24.88>cmttb10}{}

\usepackage{booktabs}
\usepackage{tabularx}
\usepackage{makecell}

\usepackage{array}
\newcolumntype{P}[1]{>{\centering\arraybackslash}p{#1}}

\newcommand{\myparagraph}[1]{\subsubsection{#1}}

\newcommand{\DAP}{DAP}
\newcommand{\DDAP}{DDAP}

\newcommand{\servers}{n}
\newcommand{\processes}{P}
\newcommand{\handler}{handler}
\newcommand{\coordhandler}{coordinator handler}
\newcommand{\msghandler}{message handler}
\newcommand{\continuation}{extension}

\makeatletter
\lst@Key{countblanklines}{true}[t]%
{\lstKV@SetIf{#1}\lst@ifcountblanklines}

\lst@AddToHook{OnEmptyLine}{%
	\lst@ifnumberblanklines\else%
	\lst@ifcountblanklines\else%
	\advance\c@lstnumber-\@ne\relax%
	\fi%
	\fi}
\makeatother

\definecolor{comment-color}{rgb}{0.6627, 0.7176, 0.7764}

\lstdefinestyle{lststyle}
{
    language = python,
    basicstyle=\footnotesize\ttfamily,
    keywordstyle = {\color{blue}\bfseries},
    morecomment=[l][\color{comment-color}]{//},
}

\author{Naama Ben-David}{VMware Research}{bendavidn@vmware.com}{}{}
\author{Gal Sela}{VMware Research, USA \and Technion, Israel}{galy@cs.technion.ac.il}{0000-0003-2342-6955}{}
\author{Adriana Szekeres}{VMware Research}{aszekeres@vmware.com}{}{}

\authorrunning{N.\, Ben-David, G.\, Sela, and A.\, Szekeres}

\Copyright{Naama Ben-David, Gal Sela, and Adriana Szekeres}



\begin{document}

\nolinenumbers




\title{The FIDS Theorems:
Tensions between Multinode and Multicore Performance in Transactional Systems}
\titlerunning{The FIDS Theorems for Multinode Multicore Transactional Systems}

\maketitle

\begin{abstract}

Traditionally, distributed and parallel transactional systems have been studied in isolation, as they targeted different applications and experienced different bottlenecks. However, modern high-bandwidth networks have made the study of systems that are both distributed (i.e., employ multiple nodes) and parallel (i.e., employ multiple cores per node) necessary to truly make use of the available hardware. 

In this paper, we study the performance of these combined systems and show that there are inherent tradeoffs between a system's ability to have fast and robust distributed communication and its ability to scale to multiple cores. 
More precisely, we formalize the notions of a \emph{fast deciding} path of communication to commit transactions quickly in good executions, and \emph{seamless fault tolerance} that allows systems to remain robust to server failures. We then show that there is an inherent tension between these two natural distributed properties and well-known multicore scalability properties in transactional systems.
Finally, we show positive results; it is possible to construct a parallel distributed transactional system if any one of the properties we study is removed.





\end{abstract}

\section{Introduction}
Transactional systems offer a clean abstraction for programmers to write concurrent code without worrying about synchronization issues. This has made them extremely popular and well studied in the last couple of decades~\cite{attiya2015disjoint,faleiro2014lazy_transactions,guerraoui2008obstruction,peluso2015disjoint,szekeres2020meerkat,yu16tictoc,zhang18tapir,tu13silo}.

Many transactional systems in practice are \emph{distributed} across multiple machines~\cite{corbett2012spanner, zhou2021foundationdb,lakshman10cassandra}, allowing them to have many benefits that elude single-machine designs. For example, distributed solutions can scale to much larger data sets, handle much larger workloads, service clients that are physically far apart, and tolerate server failures.
It is therefore unsurprising that distributed transactional systems have garnered a lot of attention in the literature, with many designs aimed at optimizing their performance in various ways: minimizing network round trips to commit transactions~\cite{kraska13mdcc,szekeres2020meerkat,zhang18tapir,mu16janus,cowling12granola,li17eris}, increasing robustness and availability when server failures occur~\cite{kraska13mdcc,zhang18tapir,szekeres2020meerkat}, and scaling to heavier workloads~\cite{zhou2021foundationdb, elhemali22dynamodb}.

Due to increased bandwidth on modern networks, new considerations must be taken into account to keep improving the performance of distributed transactional systems. In particular, while traditional network communication costs formed the main bottleneck for many applications, 
sequential processing within each node is now no longer enough to handle the throughput that modern networks can deliver (through e.g., high-bandwidth links, multicore NICs, RDMA, kernel bypassing).
Thus, to keep up with the capabilities of modern hardware, distributed transactional systems must make use of the parallelism available on each server that they use. That is,
they must be designed while optimizing \emph{both} network communication \emph{and} multicore scalability.

Two main approaches have been employed by transactional storage systems to take advantage of the multicore architecture of their servers~\cite{stonebraker1985}: \emph{shared-nothing} or \emph{shared-memory}. The shared-nothing approach, where each core can access a distinct partition of the database and only communicates with other cores through message passing, has a significant drawback: cores responsible for hot data items become a throughput bottleneck while other cores are underutilized. 
To be able to adapt to workloads that stress a few hot data items, the shared memory approach, where each core can access any part of the memory, can be used. However, shared memory must be designed with care, as synchronization overheads can hinder scalability.
Fortunately, decades of work has studied how to scale transactional systems in a multicore shared-memory setup~\cite{attiya2011inherent,attiya2012cost,attiya2015disjoint,bernstein86cc,bushkov2014pcl,papadimitriou1979serializability,peluso2015disjoint, avni08noglobalclock}.
Thus, there is a lot of knowledge to draw from when designing distributed transactional systems that also employ parallelism within each server via the shared-memory approach. 

In this paper, we study such systems, which we call \emph{parallel distributed transactional systems (PDTSs)}. Our main contribution is to show that there is an inherent tension between properties known to improve performance in distributed settings and those known to improve performance in parallel settings. To show this result, we first formalize a model that combines both shared memory and message passing systems. While such a model has been formulated in the past~\cite{aguilera2018passing}, it has not been formulated in the context of transactional systems. 

We then describe and formalize three properties of distributed transactional systems that improve their performance. These properties have all appeared in various forms intuitively in the literature~\cite{zhang18tapir, kraska13mdcc, szekeres2020meerkat}, but have never been formalized until now. We believe that each of them may be of independent interest, as they capture notions that apply to many existing systems. In particular, we first present \emph{distributed disjoint-access parallelism}, a property inspired by its counterpart for multicore systems, but which captures scalability across different distributed shards of data. Then, we describe a property that intuitively requires a \emph{fast path} for transactions: transactions must terminate quickly in executions in which they do not encounter asynchrony, failures, or conflicts. While many fast-path properties have been formulated in the literature for consensus algorithms, transactions are more complex since different transactions may require a different number of network round trips, or message delays, in order to even know what data they should access. We capture this variability in a property we name \emph{fast decision}, intuitively requiring that once the data set of a transaction is known, it must reach a decision within one network round trip. Finally, we present a property called \emph{seamless fault tolerance}, which requires an algorithm to be able to tolerate some failures without affecting the latency of ongoing transactions. This has been the goal of many recent works which focus on robustness and high availability~\cite{szekeres2020meerkat,moraru2012egalitarian,mu16janus, kraska13mdcc,zhang18tapir}.

Equipped with these properties, we then show the inherent tension that exists between them and the well-known multicore properties of disjoint-access parallelism and invisible reads, both of which intuitively improve cache coherence and have been shown to increase scalability in transactional systems~\cite{roy2009exploring,felber2010time}. More precisely, we present the \textbf{FIDS} theorem for \emph{sharded} PDTSs: a PDTS that guarantees a minimal progress condition and shards data across multiple nodes cannot simultaneously provide \textbf{F}ast decision, \textbf{I}nvisible reads, distributed \textbf{D}isjoint-access parallelism, and \textbf{S}erializability. An important implication of this result is that serializable shared-memory sharded PDTSs that want to provide multicore scalability cannot simply use a two-phase atomic commitment protocol (such as the popular two-phase commit). Furthermore, we turn our attention to \emph{replicated} PDTSs. We discover that a similar tension exists for PDTSs that utilize \emph{client-driven} replication. With client-driven replication replicas do not need to communicate with each other to process transactions. It is commonly used in conjunction with a leaderless replication algorithm to save two message delays~\cite{kraska13mdcc, zhang18tapir, szekeres2020meerkat, mu16janus}, as well as in RDMA-based PDTSs which try to bypass the replicas' CPUs~\cite{drago15nocompromise, shamis19farmv2}. We present a \emph{robust} version of the FIDS theorem, which we call the \textbf{R-FIDS} theorem: a PDTS (that may or may not shard its data) and utilizes client-driven replication 
cannot simultaneously provide \textbf{R}obustness to failures in the form of seamless fault tolerance, \textbf{F}ast decision, \textbf{I}nvisible reads, \textbf{D}isjoint-access parallelism, and \textbf{S}erializability.

Interestingly, similar impossibility proofs appear in the literature, often showing properties of parallel transactional systems that cannot be simultaneously achieved~\cite{peluso2015disjoint,attiya2011inherent,bushkov2014pcl}. Indeed, some works have specifically considered disjoint-access parallelism and invisible reads, and shown that they cannot be achieved simultaneously with strong progress conditions~\cite{attiya2011inherent,peluso2015disjoint}. However, several systems achieve both disjoint-access parallelism and invisible reads with weak progress conditions such as the one we require~\cite{yu16tictoc,tu13silo,drago15nocompromise}. To the best of our knowledge, the two versions of the FIDS theorem are the first to relate multicore scalability properties to multinode scalability ones.

Finally, we show that the FIDS theorems are minimal in the sense that giving up any one of these properties does allow for implementations that satisfy the rest.

In summary, our contributions are as follows.

\begin{itemize}
    \item We present a transactional model that combines the distributed and parallel settings.
    \item We formalize three distributed performance properties that have appeared in intuitive forms in the literature.
    \item We present the FIDS and R-FIDS theorems for parallel distributed transactional systems, showing that there are inherent tensions between multicore and multinode scalability properties.
    \item We show that giving up any one of the properties in the theorems does allow designing implementations that satisfy the rest.
\end{itemize}

The rest of the paper is organized as follows. \Cref{sec:model} presents the model and some preliminary notions. In~\Cref{sec:distributedProperties}, we define the properties of distributed transactional systems that we focus on. We present our impossibility results in \Cref{sec:impossibilities}, and then in \Cref{sec:possibilities}, we show that it is possible to build a PDTS that sacrifices any one of the properties. Finally, we discuss related works in \Cref{sec:related} and future research directions in \Cref{sec:conclusion}.

\section{Model and Preliminaries}\label{sec:model}

\begin{figure}[t!]
    \centering
    \includegraphics[scale=0.18]{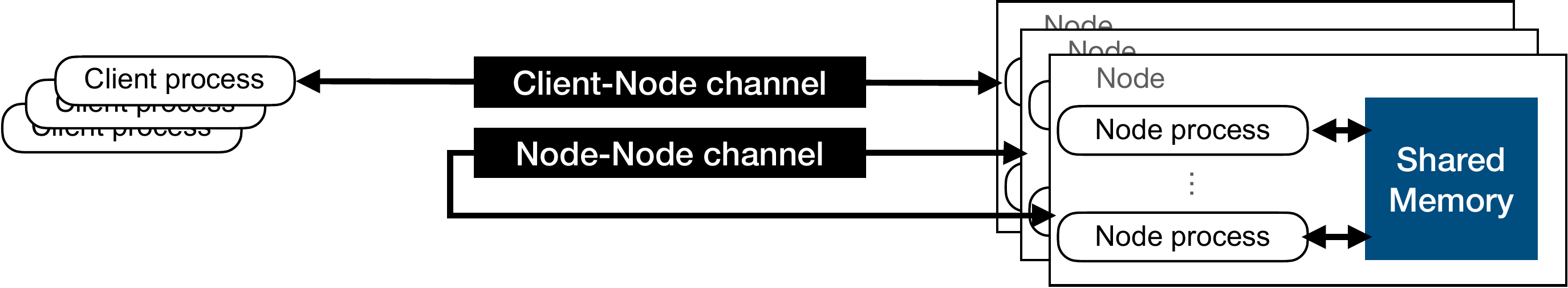}
    \caption{Communication mediums between the different types of processes considered in our model. }
    \label{fig:model-processes}
    \vspace{-4mm}
\end{figure}

\smallskip\noindent
\textbf{Communication.}
We consider a \emph{message-passing} model among $\servers$ \emph{nodes} (server hosts) and any number of \emph{client processes}, as illustrated in \Cref{fig:model-processes}. Each node has $\processes$ \emph{node processes}.
Messages are sent either between two nodes or between clients and nodes. We consider \emph{partial synchrony}~\cite{dwork1988consensus}; messages can be arbitrarily delayed
until an a priori unknown \emph{global stabilization time (GST)}, after which all messages reach their target within a known delay~$\Delta$. 
An execution is said to be \emph{synchronous} if GST is at the beginning of the execution.
Furthermore, node processes within a single node communicate with each other via \emph{shared memory}.
That is, they access \emph{shared base objects} through \emph{primitive operations}, which are atomic operations, such as read, write, read-modify-write (compare-and-swap, test-and-set, fetch-and-increment, etc.), defined in the usual way.
A primitive operation is said to be \emph{non-trivial} if it may modify the object. Two primitive operations \emph{contend} if they access the same object and at least one of them is non-trivial.
The order of accesses of processes to memory is governed by a \emph{fair scheduler} which ensures that all processes take steps.

\smallskip\noindent
\textbf{Transactions.}
We consider 
a database composed of a set of \emph{data items}, $\Sigma$, which can be accessed by \emph{read} and \emph{write} operations. Each node $N_i$ holds some subset $\Sigma_i \subseteq \Sigma$, which may overlap with the subsets held on other nodes. A \emph{transaction} $T$ is a program that executes read and write operations on a subset of the data items, called its \emph{data set}, $D_T \subseteq \Sigma$.
A transaction $T$'s \emph{write set}, $W_T \subseteq D_T$, and \emph{read set}, $R_T \subseteq D_T$, are the sets of data items that it writes and reads, respectively.
%
Two transactions are said to \emph{conflict} if their data sets intersect at an item that is in the write set of at least one of them.

\smallskip\noindent
\textbf{Transaction Interface.}
An \emph{application} may execute a transaction $T$ by calling an \emph{invokeTxn(T)} procedure.
The invokeTxn($T$) procedure returns with a \emph{commit} or \emph{abort} value indicating whether it committed or aborted, as well as the full read and write sets of $T$, with the order of execution of the operations (relative to each other), and with the read and written values. 
We say that a transaction is \emph{decided} when invokeTxn($T$) returns.


\smallskip\noindent
\textbf{Failure Model.} Nodes can fail by crashing; if a node crashes then \emph{all} processes on the node crash as well. We do not consider failures where individual processes crash and we assume clients do not fail. We denote by \emph{failure-free} execution an execution without node crashes. 

\smallskip\noindent
\textbf{Handlers and Implementations.} An \emph{implementation} of a PDTS provides data representation for transactions and data items, and algorithms for two types of \handler{}s: the \emph{\coordhandler} and the \emph{\msghandler}. Each \handler~is associated with a transaction and is executed by a single process.  Each process executes at most one \handler{} at any given time, and is otherwise \emph{idle}.
The \coordhandler{} of a transaction $T$ is the first \handler{} associated with $T$ and is triggered by an invokeTxn($T$) call 
on some client process.

The execution of a \handler~involves a sequence of \emph{handler steps}, which are of one of three types: (1) an \emph{invocation} or \emph{response} step, which is the first or last step of the handler respectively, (2) 
a primitive operation on a base object in shared memory, including its return value,
and (3) sending or receiving a message, denoted send($T$, $m$) or receive($T$, $m$). Each handler step is associated with the corresponding transaction and the process that runs it.
The return value in a response step of a transaction's \coordhandler~is the return value of invokeTxn described above, and a \msghandler~has no return value.

\smallskip\noindent
\textbf{Executions.}
An \emph{execution} of a PDTS implementation is a sequence of handler steps and \emph{node crash steps}. Each node crash step is associated with a node. 
After a node crash step associated with node $N_i$ in execution $E$, no process on node $N_i$ takes any steps in $E$. 
An execution can interleave handler steps associated with different transactions and processes.
An \emph{\continuation} $E'$ of $E$ is an execution that has $E$ as its prefix.

We say that a transaction $T$'s \emph{interval} in an execution $E$ begins at the invocation step of $T$'s \coordhandler, and ends when there are no sends associated with $T$ that have not been received whose target node has not crashed, and all \handler s associated with $T$ have reached their response step. 
 Note that the end of a transaction's interval must therefore be a response step 
of some \handler{} associated with $T$, but might not be the response step of $T$'s \coordhandler~(which may terminate earlier than some other handlers of $T$).
We say that two transactions are \emph{concurrent} in $E$ if their intervals overlap. We say that two transactions, $T_1$ and $T_2$, \emph{contend on node $N_i$} in $E$ if they are concurrent, and there is at least one primitive operation step on node $N_i$ in $E$ associated with $T_1$ that contends with a primitive operation step in $E$ associated with $T_2$. We say that $T_1$ and $T_2$ \emph{contend} in $E$ if there is some node $N_i$ such that they contend on node $N_i$ in $E$.

The \emph{projection} of an execution $E$ on a process $p$, denoted $E|p$, is the subexecution of $E$ that includes exactly all of the steps associated with $p$ in $E$. 
Two executions $E$ and $E'$ are \emph{indistinguishable} to a process $p$ if the projections of $E$ and $E'$ on $p$ are identical (i.e., if $E|p = E'|p$).

It is also useful to discuss \emph{knowledge} of properties during an execution. The notion of knowledge has been extensively used in other works~\cite{halpern1990knowledge,fagin2004reasoning}. Formally, a process $p$ \emph{knows} a property $P$ in an execution $E$ of a PDTS implementation $I$, if there is no execution $E'$ of $I$ that is indistinguishable to $p$ from $E$ in which $P$ is not true.

We adopt two concepts introduced by Lamport~\cite{lamport04lower, lamport78happenedbefore} to aid reasoning about distributed systems: \emph{depth of a step}, 
and the \emph{happened-before} relation.
The \emph{depth of a step} $s$ associated with transaction $T$ in execution $E$ is $0$ if $s$ is the invocation of $T$'s \coordhandler{}. Otherwise, it equals the maximum of (i) the depths of all steps that are before $s$ in $E$ within the same handler as $s$, and (ii) if $s$ is a receive($T$, $m$) step of a message sent in a send($T$, $m$) step, $s'$, then 1 plus the depth of $s'$.
\emph{Happened-before} is the smallest relation on the set of steps of an execution $E$ satisfying the following three conditions: 1) if $a$ and $b$ are steps of the same handler and $a$ comes before $b$ in $E$, then $a$ \emph{happened-before} $b$; 2) if $a$ is a send($T$, $m$) step and $b$ is a receive($T$, $m$) step, then $a$ \emph{happened-before} $b$; 3) if $a$ \emph{happened-before} $b$ and $b$ \emph{happened-before} $c$, then $a$ \emph{happened-before} $c$.

\smallskip\noindent
\textbf{Serializability.}
Intuitively, a transactional system is \emph{serializable} if transactions appear to have executed in some serial order~\cite{papadimitriou1979serializability}. The formal definition appears in Appendix~\ref{appendix:serializability}.

\smallskip\noindent
\textbf{Weak Progress.} A transactional system must guarantee at least \emph{weak progress}: every transaction is eventually \emph{decided}, and every transaction that did not execute concurrently with any other transaction eventually \emph{commits}.


\vspace{-2mm}

\subsection{Multicore scalability properties}
\label{sec:parallelProperties}

To scale to many processes on each server node, transactional systems should reduce memory contention between different transactions. This topic has been extensively studied in the literature on parallel transactional systems~\cite{attiya2011inherent,attiya2012cost,attiya2015disjoint,boyd2014multiprocessorbottlenecks,clements13scalablecommutativityrule,guerraoui2008obstruction,israeli94dap,peluso2015disjoint}. Here, we focus on two well-known properties, disjoint-access parallelism and invisible reads, that are known to reduce contention and improve scalability in parallel systems. We later show how they interact with distributed scalability properties.
 
\myparagraph{Disjoint-Access Parallelism}
Originally introduced to describe the degree of parallelism of implementations of shared memory primitives~\cite{israeli94dap}, and later adapted to transactional memory, \emph{disjoint-access parallelism} intuitively means 
that transactions that are disjoint at a high level, e.g., whose data sets do not intersect, do not contend on shared memory accesses~\cite{attiya2011inherent, peluso2015disjoint}. While this property may sound intuitive, it can in fact be difficult to achieve, as it forbids the use of global locks or other global synchronization mechanisms. 
Multiple versions of disjoint-access parallelism exist in the literature, differing in which transactions are considered to be disjoint at a high level. In this paper, we use the following definition. 

\begin{definition}[Disjoint-access parallelism (DAP)]
An implementation of a PDTS satisfies \emph{disjoint-access parallelism (DAP)} if two transactions whose data sets do not intersect cannot contend.
\end{definition}

\myparagraph{Invisible Reads} The second property we consider, \emph{invisible reads}, intuitively requires that transactions' read operations not execute any shared memory writes. This property greatly benefits workloads with read hotspots, by dramatically reducing cache coherence traffic.
Two variants of this property are common in the literature. The first, which we call \emph{weak invisible reads}, only requires invisible reads at the granularity of transactions. That is, if a transaction is read-only (i.e., its write set is empty), then it may not make any changes to the shared memory. This simple property has been often used in the literature~\cite{attiya2011inherent, peluso2015disjoint}.
\begin{definition}[Weak invisible reads]
An implementation of a PDTS satisfies \emph{weak invisible reads} if, in all its executions, every transaction with an empty write set does not execute any non-trivial primitives.
\end{definition}

However, this property is quite weak, as it says nothing about the number of shared memory writes a transaction may execute once it has even a single item in its write set. When developing systems that decrease coherence traffic, this is often not enough. Indeed, papers that refer to invisible reads in the systems literature~\cite{szekeres2020meerkat, tu13silo} require that no read operation in the transaction be the cause of shared memory modifications. Note that an algorithm that locally stores the read set for validation (which is the case in the above referenced systems) can still satisfy invisible reads, since the writes are not to shared memory.
Attiya et al.~\cite{attiya2012cost} formalize this stronger notion of invisible reads by requiring that we be able take an execution $E$ and replace any transaction $T$ in $E$ with a transaction that has the same write set but an empty read set, and arrive at an execution that is indistinguishable from $E$. Intuitively, this captures the requirement that reads should not update shared metadata (e.g., through ``read locks'').
We adopt Attiya et al.'s definition of invisible reads here, adapted to fit our model.

\begin{definition} [Invisible reads (due to ~\cite{attiya2012cost})]
An implementation $I$ of a PDTS satisfies the \emph{invisible reads} property if it satisfies weak invisible reads and, additionally, 
for any execution $E$ of $I$ that includes a transaction $T$ with write set $W$ and read set $R$, there exists an execution $E'$ of $I$ identical to $E$ except that it has no steps of $T$ and it includes steps of a transaction $T'$, which has the same interval as $T$ (i.e., $T$'s first and last steps in $E$ are replaced by $T'$'s ones in $E'$), and writes the same values to $W$ in the same order as in $T$, but has an empty read set. 
\end{definition}

Note that the invisible reads property complements the \DAP{} property for enhanced multicore scalability. A system that has both allows all transactions that do not conflict, not just the disjoint-access ones, to proceed independently, with no contention (as we will show in \Cref{lem:inv-dap}). Interestingly, previous works discovered some inherent tradeoffs of such systems~\cite{attiya2011inherent, peluso2015disjoint}, in conjunction with strong progress guarantees. In this paper, we study these properties under a very weak notion of progress, but with added requirements on distributed scalability (see \Cref{sec:distributedProperties}).

\section{Multinode performance properties}
\label{sec:distributedProperties}

To overcome the limitations of a single machine (e.g., limited resources, lack of fault tolerance), distributed transactional systems shard or replicate the data items on multiple nodes, and, thus, must incorporate distributed algorithms that coordinate among multiple nodes. The performance of these distributed algorithms largely depends on the number of communication rounds required to execute a transaction. Ideally, at least in the absence of conflicts, transactions can be executed in few rounds of communication, even if some nodes experience failures. In this section we propose formal definitions for a few multinode performance properties.

\subsection{Distributed disjoint-access parallelism} We start by proposing an extension of DAP to distributed algorithms, which we term \emph{distributed-DAP}, or DDAP. In addition to requiring DAP, DDAP proscribes transactions from contending on a node unless they access common elements that reside at that node:

\begin{definition}[Distributed disjoint-access parallelism (DDAP)]
    An implementation of a PDTS satisfies \emph{distributed disjoint-access parallelism (DDAP)} if for any two transactions $T$ and $T'$, and any node $N_i$, if $T$ and $T'$'s data sets do not intersect on node $N_i$ (i.e., $D_T \cap D_{T'} \cap \Sigma_i = \emptyset$), then they do not contend on node~$N_i$.
\end{definition}

While the main goal of sharding is to distribute the workload across nodes, DDAP links sharding to increased parallelism -- DDAP systems can offer more node parallelism than DAP systems through sharding.

\subsection{Fast decision}
Distributed transactional systems must integrate agreement protocols (such as \emph{atomic commitment} and \emph{consensus}) to ensure consistency across all nodes involved in transaction processing. Fast variants of such protocols can reach agreement in two message delays in ``good'' executions~\cite{lamport06lowerbounds}. Ideally, we would like distributed transactional systems to preserve this best-case lower bound, and decide transactions in two message delays; reducing the number of message delays required to process transactions not only can significantly reduce the latency as perceived by the application (processing delay within a machine is usually smaller than the delay on the network), but can also reduce the \emph{contention footprint}~\cite{faleiro2014lazy_transactions} of the transactions (intuitively, this is the duration of time in which a transaction might interfere with other transactions in the system).

Requiring transactions to be decided in just two message delays, however, is too restrictive in many scenarios. The latency of a distributed transactional system depends on how many message delays are required for a transaction to ``learn'' its data set (data items and their values); the data set needs to be returned to the application when the transaction commits, and is also used to determine whether the transaction can commit. For example, for interactive transactions or disaggregated storage, the values must be made available to the application (which runs in a client process) before the transaction can continue to execute. Thus, since the data items are remote, each read operation results in two message delays, one to request the data from the remote node and one for the remote node to reply. For non-interactive transactions or systems where transaction execution can be offloaded to the node processes, the latency for learning the data set can be improved; since the client does not need to immediately know the return value of read operations, the values of data items can be learned through a chain of messages that continue transaction processing at the nodes containing the remote data. More precisely, the client first determines a node, $n_1$, that contains the first data item the transaction needs to read; the client sends a message to $n_1$ containing the transaction; $n_1$ processes the transaction, preforming the read locally, until it determines that the transaction needs to perform a remote read from another node, $n_2$; $n_1$ sends a message to $n_2$ containing the transaction and its state so far; $n_2$ continues processing the transaction, performing the read locally, and so on. RPC chains~\cite{song09rpcchanins} already provides an implementation of this mechanism, saving one message delay per remote read operation. At the lowest extreme, non-sharded transactional systems can learn a transaction's entire data set in a single message delay. 

We introduce the \emph{fast decision} property to describe distributed transactional systems that can decide each transaction in ``good'' executions within only two message delays in addition to the message delays it requires to ``learn'' the transaction's entire data set. As explained above, the number of message delays required to learn a transaction's data set depends on several design choices. 
We note that often, deciding a transaction's outcome within two message delays after learning its data set is not plausible if the execution has suboptimal conditions, for example, if there are transactional conflicts that need to be resolved, or if not all nodes reply to messages within some timeout. This is true even for just consensus, where the two-message-delay decisions can happen only in favorable executions, on a \emph{fast path}~\cite{aguilera2020microsecond,lamport06fastpaxos}. We therefore define the fast decision property to only be required in such favorable executions.

To formalize fast decisions, we must be able to discuss several intuitive concepts more formally. In particular, we begin by defining the \emph{depth of a transaction}, to allow us to formally discuss the number of message delays that the transactional system requires to decide a transaction.

\begin{definition}[Depth of a transaction]
The \emph{depth of a decided transaction} $T$ in execution $E$ of a PDTS implementation, $d_E(T)$, is the depth of the response step of $T$'s \coordhandler{} in $E$.
\end{definition}

In many cases, we need to refer to the depth of a transaction $T$ in an execution in which $T$ is still ongoing, and its \coordhandler{} has not reached its response step yet. While we could simply refer to the depth of the deepest step of $T$ in the execution, this would not be appropriate: it is possible that a transaction in fact took steps along one `causal path' that led to a large depth, but when the response step to $T$'s \coordhandler{} happens, its depth is actually shorter. In such a case, we really only care about the depth along the `causal paths' that lead to the response step, since these are the ones affecting the latency to the application. To capture this notion, we define the \emph{partial depth} of a transaction $T$ in a prefix of an execution in which $T$ is decided as follows.

\begin{definition}[Partial depth of a transaction]
Let $T$ be a decided transaction in execution $E$ of a PDTS implementation.
The \emph{partial depth} of $T$ in a prefix $P$ of $E$ in which $T$ is not decided, $d_E(T,P)$, is the maximum step depth across all steps associated with $T$ in $P$, which happened-before the response step of $T$'s \coordhandler{} in $E$ (or $0$ if there are no such steps).
\end{definition}

We next formalize another useful concept that we need for the discussion of fast decisions; namely, what it means to \emph{learn} the data set of a transaction. For that purpose, we introduce the following two definitions:

\begin{definition}[Decided data item]
A data item $d$ is \emph{decided} to be in a transaction $T$'s read or write set in execution $E$ of a PDTS implementation if, in all \continuation{s} of $E$,
the read or write set respectively in the return value of invokeTxn($T$) contains $d$.
\end{definition}

\begin{definition}[Decided value]
A data item $d$'s \emph{value} is \emph{decided} for $T$ in execution $E$ of a PDTS implementation if, in all \continuation{s} of $E$, the read set in the return value of invokeTxn($T$) contains $d$ and with the same value.
\end{definition}

Note that a data item's value can be decided for a transaction only if that data item is part of its read set; the definition does not apply for data items in the write set. In the definition of the fast decision property and in the proofs, we refer in most places to knowing the decided values and not the data items in the write set as well. This is because knowing the read set and its values implies that a transaction's write set is decided in case it commits; this is the property that matters in many of the arguments we use in the paper.

Finally, we are ready to discuss the \emph{fast decision} property.
Intuitively, the formal definition of the property considers \emph{favorable} executions, which are synchronous, failure-free and have each transaction run solo. For those executions, the property requires two things to hold: first, a transaction is not allowed to spend more than two message delays without learning some new value for its data set, and second, once its entire data set is known, it must be decided within 2 more message delays (\Cref{cor:fast-deciding}). This captures `speed' in both learning the data set and deciding the transaction outcome.  
As discussed above, 2 message delays is an upper bound on the minimal amount of time needed to perform a read operation (and bring its value to the necessary process). Note that this is a tight bound for systems processing interactive transactions, and as such, fast decision also means optimal latency for these systems.

\begin{definition}[Fast decision]
A PDTS implementation $I$ is \emph{fast deciding} if, for every failure-free synchronous execution $E$ of $I$ and every decided transaction $T$ in $E$ that did not execute concurrently with any other transaction, for any prefix $P$ of $E$ such that $d_E(T,P)<d_E(T)-2$, there exists a prefix of $E$ of partial depth $d_E(T,P)+2$ in which the number of values known by some process to be decided is bigger than in $P$.
\end{definition}



Formalizing the allowed depth of a transaction in terms of prefixes of an execution in which the transaction is already decided (so we know its depth in that execution) helps capture the two requirements we want: (1) for any prefix of the execution, if we advance from it by two message delays, we must have improved our knowledge of the values of the read set, and (2) once the read set and its values are completely known (regardless of the depth of the prefix in which this occurs), we must be at most 2 message delays from deciding that transaction. \Cref{cor:fast-deciding} helps make this intuition concrete.

\begin{corollary}\label{cor:fast-deciding}
For every failure-free synchronous execution $E$ of a fast-deciding PDTS implementation $I$ and every decided transaction $T$ in $E$ that did not execute concurrently with any other transaction,
let $P$ be the shortest prefix of $E$ in which the value of each item in $T$'s read set is known by some process to be decided. 
 Then
\[d_E(T)\leq d_E(T,P)+2.\]
(Intuitively, $T$ must be decided within at most 2 message delays from when $T$'s read set including its values are known to be decided.)
\end{corollary}
\begin{proof}
Assume by contradiction that $d_E(T)>d_E(T,P)+2$. Then by the fast decision property of $I$, there exists a prefix of $E$ in which the number of data items whose value is known by some process to be decided is bigger than in $P$. But this is impossible, since the values of $T$'s entire read set are known to be decided in $P$.
\end{proof}

Several fast-deciding distributed transactional systems have been recently proposed for general interactive transactions~\cite{kraska13mdcc, zhang18tapir, szekeres2020meerkat}; our fast decision property captures what they informally refer to as ``one round-trip commitment''. These systems use an \emph{optimistic concurrency control} and start with an execution phase that constructs their data sets with two message delays per read operation. The agreement phase consists of validation checks that require a single round-trip latency (integrates atomic commitment and a fast consensus path in one single round trip). The write phase happens asynchronously, after the response of the transaction has been emitted to the application.

\subsection{Seamless fault tolerance}
High availability is critical for transactional storage systems, as many of their applications expect their data to be always accessible. In other words, the system must mask server failures and network slowdowns. To achieve this, many systems in practice are designed to be fault tolerant; the system can continue to operate despite the failures of some of its nodes.

However, oftentimes, while the system can continue to function when failures occur, it experiences periods of unavailability, or its performance degrades by multiple orders of magnitude while recovering~\cite{aguilera2020microsecond,wang2017apus}. This is the case in systems that must manually reconfigure upon failures~\cite{van2004chain}, and those that rely on a leader~\cite{aguilera2020microsecond,ongaro2014search,wang2017apus,lamport2001paxos,lamport1998part}.

These slow failure-recovery mechanisms, while providing some form of guaranteed availability, may not be sufficient for systems in which high availability is truly critical; suffering from long periods of severe slowdowns potentially from a single server failure may not be acceptable in some applications. 

To address this issue, some works in recent years have focused on designing algorithms that experience minimal slowdowns, or no slowdowns at all, upon failures. One approach has been to minimize the impact of leader failures by making the leader-change mechanism light-weight and switching leaders even when failures do not occur~\cite{yin2019hotstuff}. 
Another approach aims to eliminate the leader completely; such algorithms are called \emph{leaderless} algorithms~\cite{antoniadis2021leaderless,szekeres2020meerkat,zhang18tapir,moraru2012egalitarian}. 
All of these approaches aim to tolerate the failure of some nodes without impacting the latency of ongoing transactions.

In this paper, we formalize this goal of tolerating failures without impacting latency into a property that we call \emph{s-seamless fault tolerance}, where $s\leq f$. In essence, s-seamless fault tolerance requires that if only up to $s$ failures occur in an execution, no slowdown is experienced. To capture this formally, we require that for any execution $E$ with up to $s-1$ crashes, it be possible to find an equivalent execution $E'$ with one more crash event, which may happen at any time after the crashes in $E$, where the depth of all transactions are the same in $E$ and $E'$. We express this in an inductive definition.

\begin{definition}[$s$-seamless fault tolerance]
Any implementation of a PDTS satisfies $0$-seamless fault tolerance.
An implementation $I$ of a PDTS satisfies $s$\emph{-seamless fault tolerance} if it satisfies $(s-1)$-seamless fault tolerance, and for any execution $E$ of $I$ with $s-1$ node crashes, for any prefix $E_P$ of $E$ that contains the $s-1$ node crashes, and any node crash event $c$ of a node that has not crashed in $E_P$, there exists an execution $E'$ of $I$ whose prefix is $E_P\cdot c$, such that
(1) stripping each of $E$ and $E'$ of all steps other than invocation and response steps of \coordhandler{}s results in the same sequence of invocation and response steps
(intuitively, the executions are equivalent), and (2) the depth of each decided transaction is the same in both executions (intuitively, $E'$ seamlessly tolerates the node crashes).
\end{definition}


While $s$-seamless fault tolerance offers the extremely desirable robustness property, it also requires that: a) no single node can be on the critical path of all transactions, and b) no single node can be solely responsible for processing a transactional task. This can be a double-edged sword; on the one hand, this eliminates the possibility of a leader bottleneck, which implies better scalability. On the other hand, it disallows certain optimizations, like reading from a single replica.

\section{Impossibility results} \label{sec:impossibilities}

Having specified some key properties which make distributed transactional systems fast and scalable, we now turn to the main result of our paper: unfortunately, there is a tension between these multinode performance properties and the single-node but multicore performance properties discussed in \Cref{sec:model}.

More specifically, we present the \textbf{FIDS} theorems. The first FIDS theorem states that no PDTS with weak progress which shards data can guarantee \textbf{F}ast decision, \textbf{I}nvisible reads, distributed \textbf{D}isjoint-access parallelism, and \textbf{S}erializability simultaneously.
This is in contrast to known systems that achieve just the multinode properties~\cite{szekeres2020meerkat,zhang18tapir,mu16janus} or just the multicore properties~\cite{yu16tictoc,tu13silo,drago15nocompromise}. Thus, the FIDS theorem truly shows tensions that arise when a transactional system is both parallel and distributed.

The FIDS theorem refers to PDTSs that shard data. That is, each node only stores part of the database items. Interestingly, the impossibility holds in this setting even without requiring any fault tolerance, and in particular, without seamless fault tolerance. We note that the FIDS theorem applies also to systems that replicate data in addition to sharding it; adding replication on top of a sharded system only makes it more complex.

For systems that maintain multiple copies of the data, we show a different version of the result. 
Note that in such systems, distribution comes from replication; several nodes, each with a copy of the entire database, are used to ensure fault tolerance. For this setting, we present the Robust-FIDS, or \textbf{R-FIDS}, theorem: a PDTS with weak progress that utilizes client-driven replication and satisfies \textbf{R}obustness to at least one failure through the seamless fault tolerance property, in addition to satisfying \textbf{F}ast decision, \textbf{I}nvisible reads, \textbf{D}isjoint-access parallelism, and \textbf{S}erializability, is also impossible to implement.

We begin by presenting an overview of the proof technique for the two versions of the FIDS theorem, and then present the detailed proof for each of them.

\subsection{Proof Overview}\label{sec:prf-overview}

Both proofs have a similar structure; we consider example transactions that form a dependency cycle, and show an execution in which all of them commit, thereby violating serializability. To argue that all transactions in our execution commit, we build the execution by merging executions in which each transaction ran solo (and therefore had to commit by weak progress), and showing that the resulting concurrent execution is indistinguishable to each transaction from its solo run.  Starting with solo executions also gives us another property that we can exploit; we define the solo executions to be synchronous and failure-free, and therefore they must be fast deciding as well. 

The key challenge in the proofs is how to construct a concurrent execution $E_{concur}$ that remains indistinguishable to all processes from the solo execution that they were a part of. 
To do so, we divide the concurrent execution into two phases; first, we let the solo executions run, in any interleaving, until right before the point in each execution at which some process learns the values of its transaction's read set. When this point is reached in each solo execution, we carefully interleave the remaining steps in a second phase of the concurrent execution. A key feature is that by the fast decision property, which each solo execution satisfies, once some process learns the read set including its values, there are at most two message delays left in each solo execution before the transaction is decided. This bounds the amount of communication we need to worry about in the second phase of the concurrent execution.

To show that $E_{concur}$ is indistinguishable from the solo runs, we look at each of the two phases separately.
The idea is to show that no process makes \emph{any} shared memory modifications in the first phase, and then show that we can interleave messages and message handlers in a way that allows each transaction to be oblivious to the other transactions for at least one more intuitive `round trip', which is all we need to reach decision according to the fast decision property.

To show that a transaction performs no shared memory modifications in the first phase of the concurrent execution we construct, we rely on the way we choose the transactions, their data sets, and when in the execution their data sets are decided; in both proofs, the transactions we choose may have empty or non-empty write sets, depending on the results of their reads. The following lemma shows that as long as a transaction's write set is not \emph{known} to be non-empty, the transaction cannot cause any modifications in a system that provides weak invisible reads. 

\begin{lemma}\label{lem:future-writes}
Let $I$ be an implementation of a PDTS that provides weak invisible reads, and let $T$ be a transaction in an execution $E$ of $I$, such that no process in $E$ knows the following proposition: $T$'s write set is non-empty in all \continuation{}s of this execution in which $T$ is decided. Then $T$ cannot cause any base object modifications in $E$.

\end{lemma}

\begin{proof}
Let $I$ be an implementation of a PDTS that satisfies weak invisible reads.
Assume by contradiction that there is a transaction $T$ in execution $E$ of $I$, such that no process in $E$ knows that $T$'s final write set is not empty in all \continuation{}s in which $T$ is decided, and a process $p$ runs a handler associated with $T$ that performs some base object modification.

Since $p$ does not know that $T$'s final write set is not empty in all \continuation{}s of the current execution in which $T$ is decided, there exists an execution indistinguishable to $p$ from $E$ that has an \continuation{} in which $T$'s final write set is empty. Let that \continuation{} be $E_{readOnly}$. 
Since $T$'s final write set in $E_{readOnly}$ is empty, then by weak invisible reads, $T$ cannot cause base object modifications in $E_{readOnly}$. Contradiction.
\end{proof}

This lemma, combined with the way we choose the transactions in our proofs, immediately implies that phase 1 of $E_{concur}$ is indistinguishable to all processes from the solo executions they are a part of.

The proofs differ somewhat in how they show that $E_{concur}$ is indistinguishable from the solo runs in the second phase.
We argue about restricted shared memory modifications through the use of the \DAP{} and invisible reads properties in the following key lemma, which intuitively shows that transactions that do not conflict do not (visibly) contend.

\begin{lemma}\label{lem:inv-dap} Let $I$ be an implementation of a PDTS that provides both \DAP{} and invisible reads, and let $T$ be a transaction in an execution $E$ of $I$, such that its final write set is $W$. Then $T$ does not cause any base object modifications visible to any concurrent transaction in $E$ whose data set does not overlap with~$W$.
\end{lemma}

\begin{proof}
Let $I$ be an implementation of a PDTS that satisfies \DAP{} and invisible reads.
Let $T$ be a transaction whose final write set in an execution $E$ of $I$ is $W$.
Assume by contradiction that there exists some transaction $T'$ concurrent with $T$ in $E$ whose data set does not overlap with $W$, but which 
sees a modification made by $T$ in $E$.
That is, there is some base object operation step $s$ of $T'$ whose return value is affected by $T$'s modification.

By invisible reads, there exists an execution $E'$ of $I$ identical to $E$ except that it includes a transaction $T_{noRead}$ in place of $T$ with the same interval, where $T_{noRead}$ has $W$ as its write set and an empty read set.
By \DAP{}, $T_{noRead}$ does not modify in $E'$ any base object accessed by any concurrent transaction whose data set does not overlap with $W$. 
In particular, $T_{noRead}$ cannot make any modifications visible to $T'$ in $E'$. Note that step $s$ must exist in $E'$, since by definition, $E'$ is identical to $E$ except in steps associated with $T$ and $T_{noRead}$. However, in $E$, $s$'s return value is affected by $T$'s modification, and in $E'$, this modification does not exist. Therefore, $E'$ cannot be an execution of $I$. Contradiction.
%
\end{proof}

To make phase 2 of $E_{concur}$ also indistinguishable from the solo executions, we schedule the remaining messages carefully. In particular, we schedule messages sent by reading transactions to each node before those sent by writing transactions, and again rely on \DAP{} and invisible reads to argue that the reading transactions' handlers will not cause changes visible to those who write afterwards. However, here the two proofs diverge. 

\myparagraph{Sharded Systems}
The proof for sharded systems uses two nodes and two transactions, each reading from a data item on one node and, if it sees the initial value, writing on the other node. The read set of one transaction is the same as the (potential) write set of the other transaction. We need to argue that the reading transaction on some node cannot cause modifications on that node that are visible to the writing transaction. However, since the write set of each transaction overlaps with the data set of the other, we cannot apply~\Cref{lem:inv-dap}. Instead, we rely on \DDAP{}, and show that with this property, the reading transaction indeed cannot be visible to the writing one on each node. We show a lemma very similar to~\Cref{lem:inv-dap} but which applies to transactions whose write set on a specific node does not overlap the data set of another transaction on that node. 

Note that this proof relies on sharding, but does not need fault tolerance. In particular, it does not make use of the seamless fault tolerance property. However, the result does apply to systems in which the data is both sharded and replicated, as those systems are even more complex than ones in which no replication is used.

\myparagraph{Replicated But Unsharded}
In the case of a non-sharded, replicated system, we choose three transactions, where the write set of one equals the read set of the next. We divide them into pairs, where within each pair, the write set of one does not overlap with the data set of the other. We can then directly use~\Cref{lem:inv-dap} to argue that the second one to be scheduled of this pair will not see changes made by the first. We exploit fault tolerance to have the third transaction's messages never reach that node.
However, here, we must be careful, since we defined the solo executions to be failure-free to guarantee fast decisions. We therefore rely on seamless fault tolerance; we show indistinguishability of the concurrent execution not from the original solo executions, but from executions of the same depth that we know exist due to seamless fault tolerance.

Interestingly, when we convert a solo execution $S$ to an execution $F$ of the same depth (but with a node failure) via the seamless fault tolerance property, we may lose its fast decision property. That is, while the new execution must have the same depth as the original ones, that does not guarantee that it will also be fast deciding, as the fast decision property does not solely refer to the length of the execution. In particular, it could be the case that in $F$, the data set of a transaction including its values is learned earlier, but then the transaction takes more than 2 message delays to be decided. This would be problematic for our proof, in which the indistinguishability relies heavily on fast decision once the data set including its values are known. To show that this cannot happen in the executions we consider, we rely on another lemma that bounds from below the depth at which any transaction in a fault tolerant system can learn the decided values of its reads.

\subsection{Sharded Transactional Systems}

We now show that serializable sharded transactional systems that provide weak progress cannot simultaneously achieve fast decision, invisible reads and \DDAP{}. That is, sharding the data across multiple nodes while achieving these properties is impossible even if we do not tolerate any failures.

For this proof, we use a version of~\Cref{lem:inv-dap} that makes use of \DDAP{} and applies to sharded systems in which transactions do not conflict on a particular shard.

\begin{lemma}\label{lem:inv-ddap}
In any implementation of a PDTS that provides both \DDAP{} and invisible reads, a transaction whose write set is $W$ does not cause any modifications on shared based objects on a node $N$ visible to
any concurrent transaction whose data set does not overlap with~$W$ on $N$.
\end{lemma}

The proof of this lemma is very similar to the proof of~\Cref{lem:inv-dap}. The only required adjustments are using \emph{\DDAP{}} instead of \DAP{}, and referring to $T'$'s data set and $T$'s write set and modification \emph{on a certain node $N$}.

\begin{theorem}[The FIDS theorem for sharded transactional systems]\label{thm:dist-impossibility}
There is no implementation of a PDTS which shards data across multiple nodes that guarantees weak progress, and simultaneously provides fast decision, invisible reads, distributed disjoint-access parallelism and serializability.
\end{theorem}

\begin{proof}
Assume by contradiction that there exists an implementation $I$ of a PDTS with all the properties in the theorem statement.

Consider a database with $2$ data items, $X_1, X_2$, partitioned on $2$ nodes, $N_1,N_2$ respectively. Consider two transactions, $T_1, T_2$, with the following data sets: $T_1$'s read set is $\{X_1\}$. Its write set is $\{X_2\}$ if its read returns the initial value of $X_1$, in which case it writes a value different from $X_2$'s initial value. Otherwise, its write set is empty. For $T_2$, its read set is $\{X_2\}$, and its write set is $\{X_1\}$ if its read returns the initial value of $X_2$, and empty otherwise. If its write set is non-empty, it writes a value different from $X_1$'s initial value. Let $T_1$ be executed by a client $C_1$ and $T_2$ be executed by a different client $C_2$.

Consider the following executions.

\textsc{Solo Executions.} We define two executions $S_1,S_2$, corresponding to $T_1,T_2$ respectively running in isolation, without the other transaction present in the execution. Both executions are synchronous and failure-free. 
Note that by weak progress, $T_i$ commits in $S_i$, and by serializability, $T_i$ returns the initial value of its read item and therefore its write set is not empty.

\textsc{Concurrent Execution.} We define an execution, $E_{concur}$, where $T_1$ and $T_2$ execute concurrently. On each node, each transaction is executed on different processes. Recall that this can happen since this is a parallel system, and the executing processes for a transaction are arbitrarily chosen among the idle processes of each node.
In $E_{concur}$, for each transaction $T_i$, we let each process that executes it run until right before it knows the decided read set and read set value of $T_i$. Let the prefix of $E_{concur}$ that includes all these steps be $P_1$.
We then let each process that handles $T_i$ run until when the next step of its handler has depth $\geq d_{S_i}(T_i)-2$. 
Next, we let all messages sent on behalf of $T_1$ to $N_1$ and not yet received reach $N_1$ and be handled before any message sent on behalf of $T_2$ to $N_1$. For node $N_2$, we let the reverse happen; messages sent on behalf of $T_2$ reach it and are handled before messages sent on behalf of $T_1$.
Finally, we resume all processes, and pause node processes that handle $T_i$ when the next step of their handler has depth $\geq d_{S_i}(T_i)$. 
As for the client of each transaction, we let any messages sent to it arrive in the same order as they did in their corresponding solo executions (we will show that it receives the same messages in $E_{concur}$).  

We now claim that execution $E_{concur}$ is indistinguishable to $C_i$ from $S_i$, and indistinguishable to each node process running $T_i$ from the prefix of $S_i$ containing all this process's steps of depth $<d_{S_i}(T_i)$. To do so, we consider the execution in two phases; the phase before the two transactions achieve knowledge of their data sets including their values (up to the end of $P_1$), and the phase afterwards.

\textsc{Phase 1 of $E_{concur}$.}
Note that for any prefix $P$ of $E_{concur}$ in which $T_i$'s read set's value is not known to be decided by some process, $T_i$'s known decided write set in $P$ is empty. Consider the longest prefix $P_{undecided_i}$ of $P_1$ in which the decided write set of $T_i$ is still empty. Note that for every process $p$, its knowledge of $T_i$'s write set in $P_{undecided_i}$ is the same as it is in $P$. Therefore, by~\Cref{lem:future-writes}, in any such prefix $P$, $T_i$ may not make any modifications to shared base objects visible to 
\emph{any} concurrent transaction.
Therefore, in phase 1 there are no modifications visible to either transaction that were not visible in the solo execution as well. Thus, by the end of phase 1, $E_{concur}$'s prefix $P_1$ is indistinguishable to both transactions from their respective solo executions.
Therefore, both transactions read the initial values of their respective read sets, and both have a non-empty write set in $E_{concur}$.

\textsc{Phase 2 of $E_{concur}$.}
To show that $E_{concur}$ remains indistinguishable from the solo executions to their respective transactions in phase 2, we rely on the order of messages that are received by the two nodes. 

First, we note that by~\Cref{lem:inv-ddap}, $T_i$ does not make base object modifications visible to $T_{(i\text{ mod }2) + 1}$ on node $N_i$, since $T_i$'s final write set is $\{X_{(i\text{ mod }2)+1}\}$, which does not intersect $T_{(i\text{ mod }2) + 1}$'s final data set on node $N_i$.

Next, note that in each solo execution $S_i$, the first process that knows the decided value of $T_i$ must be on node $N_i$, since that is where the data for the read of $T_i$ is stored. Furthermore, by construction of $E_{concur}$, any messages sent on behalf of $T_i$ to $N_i$ immediately after both transactions gain knowledge of their write sets arrives before any such message sent on behalf of $T_{(i\text{ mod }2) + 1}$, and its handler is completely executed. Thus, by the above claim, on both nodes, all handlers of both transactions for messages sent at depth $d_{E_{concur}}(T_i, P_1)$ execute to completion in a way that is indistinguishable to $T_i$ from the solo execution $S_i$.

Finally, note that since $S_i$ is synchronous and failure and conflict free, and $I$ satisfies the fast decision property, by~\Cref{cor:fast-deciding}, the depth of $T_i$ in $S_i$ is at most 2 more than the partial depth of the first prefix in which $T_i$'s data set including its values became known. In particular, since $S_i$ is indistinguishable to processes executing $T_i$ from $E_{concur}$ up to that point, this means that $d_{S_i}(T_i) \leq d_{E_{concur}}(T_i, P_1) + 2$. Thus, once messages from within the handlers that were activated by messages sent in $E_{concur}$ at depth $d_{E_{concur}}(T_i, P_1)$ are received, $T_i$ must be decided in $E_{concur}$ as well, since $E_{concur}$ is indistinguishable from $S_i$ to all processes running $T_i$ up to this point. Therefore, both transactions commit successfully in $E_{concur}$ in a manner indistinguishable from their respective solo executions. 

However, this means that there is a circular dependency between the two transactions; $T_2$ must occur before $T_1$, since it returns the initial value of $X_2$, before $T_1$ writes to it. Similarly, $T_1$ must occur before $T_2$, since it returns the initial value of $X_1$. This therefore contradicts serializability.
\end{proof}

\subsection{Replicated Unsharded Systems}
So far, we have considered a PDTS in which node failures cannot be tolerated; if one of the nodes crashes, we lose all data items stored on that node, and cannot execute any transactions that access those data items.
However, in reality, server failures are common, and therefore many practical systems use replication to avoid system failures. Of course, the impossibility result of Theorem~\ref{thm:dist-impossibility} holds for a PDTS even for the more difficult case in which failures are possible and each node's data is replicated on several backups.

However, we now turn our attention to PDTSs in which the entire database is stored on each node. This setting makes it plausible that a client could get away with accessing only one node to see the state of the data items of its transaction. However, we show that the impossibility of Theorem~\ref{thm:dist-impossibility} still holds for a system in which failures are tolerated without affecting transaction latency (i.e., systems that satisfy seamless fault tolerance).

As explained in~\Cref{sec:prf-overview}, the use of seamless fault tolerance requires us to explicitly argue about the length of the executions in which transactions decide. To do so, we need the following lemma, while gives a lower bound for the depth at which a transaction's read set and values can be decided.

\begin{lemma}\label{lem:read-delay}
There is no execution $E$ of any 
serializable PDTS implementation that tolerates at least 1 failure in which there is a transaction $T$ and prefix $P$ such that $d_E(T, P) < 2$ and some process knows the decided value of some read of $T$ in $P$.
\end{lemma}

\begin{proof}
Assume by contradiction that there is some implementation $I$ of a serializable PDTS that tolerates at least 1 failure, an execution $E$ of $I$, and a prefix $P$ of $E$ such that $d_{E}(T,P) \leq 1$ and some process knows the decided value of some data item $d$ of $T$ in $P$. Without loss of generality, let process $p$ on node $N$ be the process that knows $d$'s decided value, let that value be $v$ and let $T$'s invoking client be $C$. Note that since $C$ does not have access to the data, and any step of any process not on $C$'s node must be of depth at least 1, $p$ cannot be on $C$'s node, and cannot have received any message from any process other than $C$ within depth less than 2. Therefore $p$ can only know the value of $d$ on node $N$, but not any other nodes.

Consider the following executions.

$E_{N-\mathit{fail}}$. $E_{N-\mathit{fail}}$ and $E$ are identical up to right before $T$'s invocation. In $E_{N-\mathit{fail}}$, node $N$ fails at this point. Then, a transaction $T'$ is invoked by a client $C' \neq C$. $T'$ writes a value $v' \neq v$ to $d$ and commits. After $T'$ commits, $T$ is invoked in $E_{N-\mathit{fail}}$. Clearly, by serializability, $T$'s read of $d$ in $E_{N-\mathit{fail}}$ returns $v'$ or a more updated value, but not $v$.

$E_{N-\mathit{slow}}$. $E_{N-\mathit{slow}}$ is identical to $E_{N-\mathit{fail}}$ except that node $N$ does not fail in $E_{N-\mathit{slow}}$. Instead, all messages to and from $N$ are arbitrarily delayed in $E_{N-\mathit{slow}}$ starting at the same point at which $N$ fails in $E_{N-\mathit{fail}}$. Clearly, $E_{N-\mathit{slow}}$ is indistinguishable from $E_{N-\mathit{fail}}$ to all processes not on $N$.

$E'$. $E'$ is identical to $E_{N-\mathit{slow}}$ except that node $N$ receives messages from client $C$. Clearly, $E'$ and $E_{N-\mathit{slow}}$ are indistinguishable to all processes not on $N$. So, $T$'s read of $d$ must return the same value as in $E_{N-\mathit{slow}}$, namely $v'$ or a more updated one, but not $v$. However, note that $E'$ is also indistinguishable to processes on $N$ from $E$ in any prefix of $E$ of partial depth $<2$ for $T$, since no process in $N$ received any messages other than those it received in $E$, and since clients do not receive any messages not related to their own transactions, so $C$ must have sent the same message(s) to $N$ in $E'$ as it did in $E$. Therefore, there is a prefix $P'$ of $E'$ indistinguishable to $p$ from $P$, in which $v$ is not the decided value of $d$, contradicting $p$'s knowledge of $d$'s decided value in $P$.
\end{proof}

\begin{theorem}[The R-FIDS theorem for replicated transactional systems]\label{thm:repl-impossibility}
There is no implementation of a PDTS that utilizes client-driven replication that guarantees weak progress, and simultaneously provides $1$-seamless fault tolerance, fast decision, invisible reads, disjoint-access parallelism and serializability.
\end{theorem}

The proof of this theorem is similar to the proof of Theorem~\ref{thm:dist-impossibility}. We build a cycle of dependencies between transactions where each neighboring pair in the cycle overlaps on a single data item that one of them reads and the other writes. The key is that because of invisible reads, each read can happen before the write on the same data item without leaving a trace. However, to construct this cycle in the replicated case, we need at least 3 replicas, 3 transactions and 3 data items. This is because we can no longer separate the read and write of a single transaction on each node.
Furthermore, we make use of~\Cref{lem:read-delay}, as well as the budgeted depth of a transaction in a fast-deciding execution, to explicitly argue about the amount of communication possible after a transaction learns its write set.
The full proof is presented in \Cref{sec:impossibilityProof}.

\section{Possibility results}\label{sec:possibilities}


Any subset of the properties outlined in \Cref{thm:repl-impossibility} \emph{is} possible to achieve simultaneously in a single system. Due to lack of space, we show this in \Cref{sec:appendix:possibilities}, where we present four distributed transactional system algorithms, each sacrificing one of the desired properties. Recall from our model description that the presented protocols work under the assumption that the client does not fail and nodes do not recover, and as such are not intended to be used ``as is'' in practice. 
We first present a `base' algorithm which achieves all the desired properties (i.e., fast decision, invisible reads, \DDAP{}, and 1-seamless fault tolerance), but is not serializable. We obtain each of the four transactional systems algorithms, by tweaking the base algorithm to sacrifice one desired property and gain serializability.

\section{Related Work}\label{sec:related}

Disjoint-access parallelism was first introduced in~\cite{israeli94dap} in the context of shared memory objects. It was later adapted to the context of transactions. Over the years, it has been extensively studied as a desirable property for scalable multicore systems~\cite{yu16tictoc,tu2013speedy,szekeres2020meerkat,attiya2011inherent,attiya2015disjoint,peluso2015disjoint,guerraoui2008obstruction}. Several versions of \DAP{} have been considered, differing in what is considered a conflict between operations (or transactions). A common variant of \DAP{} considers two transactions to conflict if they are connected in the conflict graph of the execution (where vertices are transactions and there is an edge between two transactions if their data sets intersect)~\cite{attiya2011inherent,peluso2015disjoint}. In this paper, we consider a stricter version, which only defines transactions as conflicting if they are neighbors in the conflict graph. This version has also appeared frequently in the literature~\cite{peluso2015disjoint,bushkov2014pcl}.

Invisible reads have also been extensively considered in the literature~\cite{scherer2005advanced,attiya2011inherent,szekeres2020meerkat,tu2013speedy,yu16tictoc,herlihy2003software}. Many papers consider invisible reads on the granularity of data item accesses; any read operation on a data item should not cause changes to shared memory~\cite{tu2013speedy,szekeres2020meerkat,herlihy2003software}. Others, often those that study invisible reads from a more theoretical lens, consider only the invisibility of \emph{read-only transactions}~\cite{attiya2011inherent,peluso2015disjoint}.

Some impossibility tradeoffs for transactional systems, similar to the one we show in this paper, are known in the literature. Attiya et al.~\cite{attiya2011inherent} show that it is impossible to achieve weak invisible reads, disjoint-access parallelism, and wait-freedom in a parallel transactional system. Peluso et al.~\cite{peluso2015disjoint} show an impossibility of a similar setting, with disjoint-access parallelism, weak invisible reads, and wait-freedom, but consider any correctness criterion that provides real-time ordering. Bushkov et al.~\cite{bushkov2014pcl} show that it is impossible to achieve disjoint-access parallelism and obstruction-freedom, even when aiming for consistency that is weaker than serializability. In this paper, none of our algorithms provide the obstruction-freedom considered in~\cite{bushkov2014pcl}; we use locks, and our algorithms can therefore indefinitely prevent progress if process failures can occur while holding locks.

Fast paths for \emph{fast decision} have been considered extensively in the replication and consensus literature~\cite{keidar2001cost,dutta2005fast, lamport06fastpaxos, aguilera2019impact, aguilera2020microsecond}. In most of these works, the conditions for remaining on the fast path include experiencing no failures. That is, they do not provide seamless fault tolerance. However, some algorithms, like Fast Paxos~\cite{lamport06fastpaxos}, can handle some failures without leaving the fast path. 
In the context of transactional systems, the fast path is often considered for conflict-free executions rather than those without failures~\cite{szekeres2020meerkat, zhang18tapir, mu16janus, kraska13mdcc}, as we do in this paper. Seamless fault tolerance captures the idea that (few) failures should not cause an execution to leave the fast path. Systems often have a general fault tolerance $f$ that is higher than the number of failures they can tolerate in a seamless manner~\cite{szekeres2020meerkat, zhang18tapir, mu16janus, kraska13mdcc}.

Seamless fault tolerance as presented in this paper is also related to \emph{leaderlessness}~\cite{moraru2012egalitarian,antoniadis2021leaderless}, as any leader-based algorithm would slow down upon a leader failure. However, the leaderless requirement alone is less strict than our seamless fault tolerance; Antoniadis et al.~\cite{antoniadis2021leaderless} defined a leaderless algorithm as any algorithm that can terminate despite an adaptive adversary that can choose which process to temporarily remove from an execution at any point in time. This does not put any requirement on the speed at which the execution must terminate.

Parallel distributed transactional systems have been recently studied in the systems literature. Meerkat~\cite{szekeres2020meerkat} provides serializability and weak progress, and three of the desirable properties we outline in this paper. It does not, however, provide invisible reads in any form (not even the weaker version). Eve~\cite{kapritsos12eve} considers replication for multicore systems. It briefly outlines how PDTS transactions are possible using its replication system, but it is not their main focus. 

\section{Discussion}\label{sec:conclusion}

This paper is inspired by recent trends in network capabilities, which motivate the study of distributed transactional systems that also take advantage of the parallelism available on each of their servers. 
We formalize three performance properties of distributed transactional systems that have appeared intuitively in various papers in the literature, and show that these properties have inherent tensions with multicore scalability properties. 
In particular, in this paper we formalized the notions of distributed disjoint-access parallelism, a fast decision path for transactions, and robustness in the form of seamless fault tolerance. Combined with the well-known multicore scalability properties of disjoint-access parallelism and invisible reads, we show the \emph{FIDS theorem}, and its fault tolerant version, the \emph{R-FIDS theorem}, which show that serializable transactional systems cannot satisfy all these properties at once. Finally, we show that removing any one of these properties allows for feasible implementations.

We note that our possibility results can be seen as ``proofs of concept'' rather than practical implementations. It would be interesting to design practical algorithms that give up just one of the properties we discuss. We believe that each property has its own merit for certain applications and workloads, and it would be interesting to determine which property would be the best to abandon for which types of applications.

In this work, we focused on studying parallel distributed transactional systems under a minimal progress guarantee. It would also be interesting to explore PDTSs under stronger progress conditions, or consistency conditions other than serializability. It would be equally interesting to see if the tension still exists between weaker variants of the properties we considered.


\newpage
\bibliography{biblio}

\appendix

\section*{Appendix}

\section{Formal Definition of Serializability}\label{appendix:serializability}

A committed history $H$ is \emph{derived} from an execution $E$ if it consists of exactly the following events: for each committed transaction $T$ in $E$, 
$H$ includes read and write events associated with $T$ for all the reads and writes in the response step of $T$'s \coordhandler{}, including the read or written value, in the order stated in the response step's return value (which is the return value of the call to invokeTxn($T$)).
The events of different transactions may be interleaved in any order in a derived committed history.
A committed history $S$ is \emph{sequential} if it contains no overlapping transactions, namely, events of different transactions do not interleave each other. $S$ is \emph{legal} if each read of a data item returns the last value written to that data item (or its initial value if no write operation was applied to it so far).
An execution of a PDTS is \emph{serializable} if it derives some legal sequential committed history.
A PDTS is serializable if all its executions are serializable.
\section{DDAP Lemma for FIDS Theorem}

\begin{lemma}[\Cref{lem:inv-ddap}.]
In any implementation of a PDTS that provides both \DDAP{} and invisible reads, a transaction whose write set is $W$ does not cause any modifications on shared based objects on a node $N$ visible to
any concurrent transaction whose data set does not overlap with~$W$ on $N$.
\end{lemma}

\begin{proof}
Let $I$ be an implementation of a PDTS that satisfies \DDAP{} and invisible reads.
Let $T$ be a transaction whose final write set on a node $N$ in an execution $E$ of $I$ is $W$.
Assume by contradiction that there exists some transaction $T'$ concurrent with $T$ in $E$ whose data set on $N$ does not overlap with $W$, but which 
sees a modification made by $T$ on a base object on $N$ in $E$.
That is, there is some base-object operation step $s$ of $T'$ whose return value is affected by $T$'s modification.

By invisible reads, there exists an execution $E'$ of $I$ identical to $E$ except that it includes a transaction $T_{noRead}$ in place of $T$ with the same interval, where $T_{noRead}$ has $W$ as its write set on $N$ and an empty read set.
By \DDAP{}, $T_{noRead}$ does not modify in $E'$ any base object on $N$ accessed by any concurrent transaction whose data set does not overlap with $W$ on $N$. 
In particular, $T_{noRead}$ cannot make any modifications visible to $T'$ on $N$ in $E'$. Note that step $s$ must exist in $E'$, since by definition, $E'$ is identical to $E$ except in steps associated with $T$ and $T_{noRead}$. However, in $E$, $s$'s return value is affected by $T$'s modification, and in $E'$, this modification does not exist. Therefore, $E'$ cannot be an execution of $I$. Contradiction.
\end{proof}

\section{R-FIDS Impossibility Proof}\label{sec:impossibilityProof}

\begin{proof}[Proof of Theorem~\ref{thm:repl-impossibility}.]
Assume by contradiction that there exists an implementation $I$ of a parallel replicated transactional system with all the properties stated in the theorem.

Consider a transactional system with 3 nodes $N_1, N_2, N_3$.  (For less than 3 nodes, there is no PDTS that tolerates $f\geq1$ failures in the partial-synchrony model~\cite{dwork1988consensus}.)
Further consider 3 transactions $T_1, T_2, T_3$, 3 client processes $C_1, C_2, C_3$, and 3 data items $X_1, X_2, X_3$ each of which is replicated on all 3 nodes. The data sets of the transactions are as follows:
$T_i$'s read set includes $X_{(i\text{ mod }3)+1}$, and if the result of $T_i$'s read of $X_{(i\text{ mod }3)+1}$ is the initial value of $X_{(i\text{ mod }3)+1}$, its write set includes $X_i$. Otherwise, its write set is empty. Each transaction $T_i$, if its write set is non-empty, writes a value that is different from $X_i$'s initial value.

\begin{center}
\begin{tabular}{ |P{0.4cm}!{\vrule width 1pt}P{2.1cm}|P{2.1cm}|P{2.1cm}|  }
 \hline
 \multicolumn{4}{|c|}{Transactions read and write sets} \\
 \Xhline{2\arrayrulewidth}
 $T$ & $T_1$ & $T_2$ & $T_3$ \\
 \Xhline{3\arrayrulewidth}
 $R_T$ & $\{X_2\}$ & $\{X_3\}$ & $\{X_1\}$ \\
 \hline
 $W_T$ & $\{X_1\}$ if R($X_2$)=$\bot$, else \{\} & $\{X_2\}$ if R($X_3$)=$\bot$, else \{\} & $\{X_3\}$ if R($X_1$)=$\bot$, else \{\} \\
 \hline
\end{tabular}
\end{center}

Consider the following executions. 
For each $i = 1,2,3$, in any of the following executions, if it includes $T_i$ then its \coordhandler{} is executed by $C_i$. 

\textsc{Solo Executions.} Let $E_1, E_2, E_3$ be failure-free synchronous executions of $I$, where transaction $T_i$ runs solo in $E_i$. 
Since $E_i$ contains a single transaction and $I$ satisfies weak progress, transaction $T_i$ commits in $E_i$.
Since $E_i$ is synchronous, has no failures and contains only $T_i$, and $I$ satisfies fast decision, $T_i$ is fast deciding in $E_i$. 

\textit{Claim: $T_i$ must have a depth of at most 4 in $E_i$.}

To see this, note that by the definition of fast decision, if transaction $T_i$ in $E_i$ has depth at least 3, the empty prefix of $E_i$ must have an \continuation{} $C_i$ of partial depth $d_{E_i}(T_i, C_i) \leq 2$ in which the value of the read set's item is known by some process to be decided, and therefore the write set is known by that process to be decided as well. By~\Cref{cor:fast-deciding}, the depth of $T_i$ in $E_i$ must be at most 2 more than the depth of $C_i$, and therefore is at most 4.

Since $I$ satisfies $1$-seamless fault tolerance, there exist executions $E'_1, E'_2, E'_3$ of $I$, where the first event in $E'_i$ is a crash of $N_i$, $T_i$ runs solo and the depth of $T_i$ in $E'_i$ is the same as its depth in $E_i$.
We assume that in each $E'_i$, a different set of processes runs the handlers.
Lastly, since $I$ is serializable, each $E'_i$ is serializable, thus $T_i$'s read in $E'_i$ returns the initial value of $X_{(i\text{ mod }3)+1}$, and therefore modifies $X_i$ as part of its write set.

Since $T_i$'s write set is only determined from the outcome of $T_i$'s read, 
and may be empty until that read's value is decided, 
by~\Cref{lem:future-writes}, no base object modifications visible to other transactions are executed by $T_i$ in $E'_i$ until after $T_i$'s read set values are known to some process. Let the shortest prefix at which some process gains knowledge of $T_i$'s read set values in $E'_i$ be $P_i$.
By~\Cref{lem:read-delay}, $d_{E'_i}(T_i, P_i) \geq 2$. 

\textsc{Concurrent Execution.} We define an execution $E_{concur}$ with all 3 transactions.
In $E_{concur}$, all messages between processes on node $N_i$ and any process that executes handlers associated with $T_i$ are arbitrarily delayed. 
For each $i = 1,2,3$, we let the processes that execute $T_i$ in $E'_i$ run in $E_{concur}$ identically to $E'_i$, in the same order of steps, until the end of $P_i$.

Note that up to this point, $E_{concur}$ is indistinguishable to all executing processes from the solo executions, since none of them has made any shared memory modifications visible to the others. Therefore, the prefix $P_{knowledge}$ of $E_{concur}$ up to this point is an execution of $I$.

We now continue $E_{concur}$ as follows: 
We let all messages sent on behalf of $T_i$ at depth $d_{E'_i}(T_i, P_i)$ be sent in $E_{concur}$, and be received and handled in the following order:
on node $N_1$, messages for $T_2$ are received first, and their handlers are run to completion, followed by messages for $T_3$. On node $N_2$, $T_3$'s messages are handled first, followed by messages of $T_1$. Finally, on node $N_3$, messages of $T_1$ are handled first followed by messages of $T_2$.
\coordhandler{s} receive messages in the same order they received them in their corresponding solo executions.

Recall that transaction $T_i$  reads data item $X_{(i\text{ mod }3)+1}$ and, if it reads the initial value, writes data item $X_i$. Thus, the service order defined above for execution $E_{concur}$ (see the order in which the nodes process their writes in \Cref{fig:repl-cycle}) means that on each node, the second serviced transaction writes to data item $X$ after the first transaction reads $X$, but it is never the case that a transaction reads a data item after it was written by another transaction on the same node. 
Since the data set of the second transaction to execute handlers after prefix $P_{knowledge}$ on each node does not overlap with the write set of the first one, and since $I$ provides invisible reads and \DAP{}, then by Lemma~\ref{lem:inv-dap}, the first transaction does not make base object modifications visible to the second transaction. In other words, on each node, a process executing the second transaction cannot observe any changes on shared memory. Thus, $E_{concur}$ is still indistinguishable from $E'_i$ to any node process that executes $T_i$ up to the end of the handlers of messages sent at depth $d_{E_{concur}}(T_i, P_i)$. 
Note, however, that since $d_{E'_i}(T_i) \leq 4$ and $d_{E_{concur}}(T_i, P_i) \geq 2$, this means that $E_{concur}$ remains indistinguishable from $E'_i$ to these processes until $T_i$ is decided. 

Therefore, for all three transactions $T_i$, $T_i$ commits successfully in $E_{concur}$, reading the initial value of $X_{(i\text{ mod }3)+1}$ and writing a non-initial value in $X_i$.


%

However, this means that there is a circular dependency between the transactions (transaction $T_1$ must happen before $T_2$, which must happen before $T_3$, which must happen before $T_1$), which contradicts serializability.
\end{proof}

\vspace{-1mm}

\begin{figure}[t!]
\begin{minipage}{0.45\linewidth}
\centering
\begin{tabular}{cccccc}
\textbf{}                 & $X_1$ & $X_2$                            & {\color[HTML]{000000} $X_3$} \\
$N_1$                     &       & 1                                & {\color[HTML]{000000} 2} \\
$N_2$                     & 2     & {\color[HTML]{303498} \textbf{}} & 1 \\
$N_3$                     & 1     & {\color[HTML]{000000} 2}
\end{tabular}
\label{tab:repl-proof}
\end{minipage}
\caption{Visual representation of execution $E_{concur}$ in the proof of Theorem~\ref{thm:repl-impossibility}. The numbers in the table represent the order of writing on each node; on node $N_1$, $X_2$ is written first, followed by $X_3$, and so on.}
    \label{fig:repl-cycle}
\vspace{-1mm}
\end{figure}

\section{Possibility results}\label{sec:appendix:possibilities}


In this section, we show that any subset of the properties outlined in Theorem~\ref{thm:repl-impossibility} \emph{is} possible to achieve simultaneously in a single system. To do so, we present four distributed transactional system algorithms, each sacrificing one of the desired properties. Recall from our model description that the presented protocols work under the assumption that the client does not fail and nodes do not recover, and as such are not intended to be used ``as is'' in practice. 

Our algorithms all have a similar structure, with minor tweaks to guarantee or sacrifice certain properties. We therefore begin by presenting a `base' algorithm. This algorithm achieves all the desired properties (i.e., fast decision, invisible reads, \DDAP{}, and seamless fault tolerance), but is not serializable. In each of the following subsections, we tweak the base algorithm to sacrifice one desirable property and gain serializability.
We note that we refer here to \DDAP{} though \Cref{thm:repl-impossibility} refers to \DAP{}, since \DDAP{} is stronger than \DAP{} (it implies \DAP{}), and so our possibility results are accordingly stronger using \DDAP{}.

The base algorithm works as follows. Each data item $d$ is replicated on $0<k<=n$ nodes, to which we refer as $d$'s \emph{replica group}. If $k < n$ then we assume the system is sharded. All nodes execute the same state machine both in the sharded and the fully replicated case (in which $k=n$). Each data item maintains its current value and a sequence number. Each data item is also assigned two locks: a short-lived lock, $lock_S$, that ensures that the value and the sequence number of the data item are read and written atomically, and a long-lived lock, $lock_L$, utilized by the distributed concurrency control mechanism. We adopt an interactive transaction model for executing transactions, where the initiating client coordinates its own transaction.
There are two phases per transaction:

\textbf{Execution phase.} During the execution phase the client reads items and dynamically constructs its data set. A process receiving an execution phase message from a client waits until the relevant data item $lock_S$ is unlocked, and then reads the data item and sends its value and sequence number to the client. To ensure that the sequence number and the data item's value are read atomically while still maintaining the invisible reads property (e.g., there are no read locks), we employ a lock-free read mechanism, as follows. To read a data item, the reader checks that its $lock_S$ is not acquired, then checks the sequence number, and then verifies that the lock is still not taken. If that verification passes, it now reads the data item's value, and finally checks the sequence number one more time. If the sequence numbers are the same and the lock was free on both checks, it succeeds.  Otherwise, it fails (the reader can retry or send an abort message to the client). Writers first acquire the lock on the data item, then change its sequence number, and only then modify the value. This mechanism is similar to the one used in Silo~\cite{tu13silo}, and guarantees atomic reads of the sequence number and value, even if they cannot be read atomically together in hardware.
During this execution phase, the client records the sequence number of each data item it reads. 


\begin{figure}[t!]
\begin{lstlisting}[style=lststyle, frame=tlrb]
//Execution phase
for each key r in the read set: 
    Send <READ,r> to all nodes that have r
    Wait to receive (val, seqNum) from k-f nodes
    Record for r the (val, seqNum) of the message with the max seqNum

Create transaction message T=<tid,sequence>, where sequence specifies the data item accesses, with reads of the form (key, seqNum) and writes of the form (key, newVal)

//Validation Phase
Send <VALIDATE,T> to all nodes
Wait for n-f responses
if all responses are of COMMIT type:
    for each key w in the write set:
        Append seqNum to w in T, where seqNum = max seqNum for w across all responses + 1
    Commit T
    Send <COMMIT,T> to all nodes
else:
    Abort T
    Send <ABORT,T> to all nodes
\end{lstlisting}
    \caption{Client code in the base algorithm.}
    \label{alg:client}
    \vspace{-1mm}
\end{figure}

\textbf{Validation phase.}
The client then executes a validation phase in which it communicates with the nodes to verify that its read values are still valid (the data items have the same sequence number) and to update the items in its write set. 
A process receiving a validation phase message on node $N_i$ iterates through the data items $d$ in the message such that $d$ is stored in node $N_i$; for each item in the read set, the process checks whether the data item $d$'s long-lived lock $lock_L$ is locked, and whether its current sequence number matches the one specified by the client. For each item in the write set, the process tries to acquire $d$'s $lock_L$.
If at any point, it runs into a data item that is already locked or its sequence number is out of date, it releases the data items it has already locked (if any), and replies `abort' to the client. Otherwise, if all data items' sequence numbers matched the client's validation phase message and (if it's not a read-only transaction) the process managed to lock them itself, the process sends `commit' to the client. 

The client waits to receive $n-f$ replies from each shard, where $f$ is the number of failures the algorithm tolerates (at most $(k-1)/2$ for $k$ replicas of each shard). If any of them was `abort', the transaction is aborted. Otherwise, the transaction is committed. The client then lets all nodes know whether the transaction committed by sending another message. Note that this message does not affect the fast decision property, since this is after the client knows the outcome of the transaction. A process receiving this final message applies the writes to the transaction's write set if it committed, and then releases all locks. 
We present pseudocode 
for this algorithm split by client code (\Cref{alg:client}) and node process code (\Cref{alg:proc}). 

\begin{figure}[t!]
\begin{lstlisting}[style=lststyle, frame=tlrb]
Upon receiving <READ,r> from C:
    Let d be the data item with key r
    // lock-free atomic read of r's seqNum and val
    wait until d.@$lock_S$@ is not locked
    int seqNum = d.@$seqNum$@
    if d.@$lock_S$@ is locked: retry
    data val = d.@$val$@
    if seqNum != d.@$seqNum$@: retry
    Send (val, seqNum) to C

Upon receiving <VALIDATE,T> from C:
    bool success = true
    List<key,seqNum> writeSeqs = []
    for each data item d corresponding to an item of T:
        if d.@$lock_L$@ != None:
            success = false; break
        if d is in the read set of T:
            if d.@$seqNum$@ != seqNum of d in T:
                success = false; break
        else:
            if not d.@$lock_L$@.CAS(None, T.tid):
                success = false; break
            Append <d.@$key$@,d.@$seqNum$@> to writeSeqs
    if success:
        Send <COMMIT,writeSeqs> to C
    else:
        Set all successfully-CASed @$lock_L$@ to None
        Send ABORT to C

Upon receiving <COMMIT,T> from C:
    for <k,newVal,seqNum> in the write set of T:
        Let d be the data item with key k
        d.@$lock_S$@.lock()
        if d.@$seqNum$@ @$<$@ seqNum:
            d.@$seqNum$@ = seqNum
            d.@$val$@ = newVal
        d.@$lock_S$@.unlock()
        if d.@$lock_L$@ == T.tid:
            d.@$lock_L$@ = None

Upon receiving <ABORT,T> from C:
    for each data item d corresponding to a write set item of T:
        if d.@$lock_L$@ == T.tid:
            d.@$lock_L$@ = None
\end{lstlisting}
\caption{Process code in the base algorithm.}
\label{alg:proc}
\vspace{-1mm}
\end{figure}

It is easy to see that this algorithm satisfies all 4 desired properties; each data item has its own associated base objects, and a transaction only accesses data item $d$ if $d$ is in its data set. Therefore, \DDAP{} holds. Furthermore, non-trivial primitives are only applied to base objects associated with a data item in a transaction's write set. In particular, only the write set is locked, and the value and sequence numbers are only updated on locked data items. Thus, the invisible reads property is satisfied; any two transactions whose write sets are the same execute non-trivial primitives on the same set of base objects.
The other two properties hold because of the way the client operates; the client only sends one message and waits for one response from $n-f$ nodes in the validation phase before it knows the outcome of the transaction. Furthermore, the execution phase causes only trivial primitives on shared memory, and therefore sends no non-trivial messages. Thus, the fast decision and $f$-seamless fault tolerance properties also hold.
However, the impossibility results apply to this algorithm, since it is not serializable (it provides the read committed isolation guarantee; the lack of serializability can be checked by running the scenarios presented in our impossibility proofs). 

\subsection{Sacrificing fast decision}\label{sec:tweak-ORT}

In this subsection we show how to achieve serializability by sacrificing the fast decision property. To do so, we split the validation phase of the client into two round trips (4 message delays), similarly to~\cite{drago15nocompromise}, but leave the rest of the algorithm the same. In this version, the client first sends a message with just its write set, and the nodes attempt to lock all data items in that set. If all nodes are successful, the client then sends its read set for the nodes to verify that the sequence numbers have not changed. If all nodes observe the same sequence number for all data items in the read set, then the client can commit. Otherwise, the transaction is aborted.
Clearly, this algorithm does not fast decide, but does not damage the other properties of the base algorithm we considered. 


\textbf{Proof of serializability.} 
We now briefly argue that the algorithm satisfies serializability.
Recall that the client first reads all items in its read set and learns their sequence number. It then acquires all write locks in the first round of the validation phase, and then rechecks the sequence numbers of read items in the second round. Note that data items change if and only if their sequence number changes as well. Thus, if no sequence number changed in the read set, then none of these items have been modified by any other process between the time the client read the last item in the execution phase and the time it read the first one in the validation phase. Thus, the transaction can serialize at the point at which it holds all write locks. 

For the replicated case, as long as $f<k/2$, any two client quorums will intersect at at least one node of each replica group of each data item common to both transactions; this guarantees that concurrent conflicting transactions cannot both commit.



\subsection{Sacrificing invisible reads}\label{sec:tweak-IR}

We now present a simple transactional memory algorithm that satisfies serializability, \DDAP{}, fast decision, $f$-seamless fault tolerance, and \emph{weak} invisible reads. That is, we show that weakening the invisible reads property suffices to make such a system feasible. 

To relax the invisible reads property, but guarantee serializability, we tweak the base algorithm to acquire more locks. In particular, instead of just acquiring the locks of data items in a transaction's write set, processes now also acquire the locks for the data items in the read set. However, to achieve the weak invisible reads property, this is only done for transactions that have a non-empty write set. The rest of the algorithm behaves exactly like the base algorithm.
This algorithm is very similar to the one presented by Szekeres et al. in Meerkat~\cite{szekeres2020meerkat}. The main difference here is that we avoid non-trivial primitives by read-only transactions, to maintain the weak invisible reads property.  

Clearly, this algorithm preserves \DDAP{}, fast decision, and $f$-seamless fault tolerance. Furthermore, by construction, it is clear that it satisfies the weak invisible reads property.

\textbf{Proof of Serializability.}
To argue about serializability, we show a reduction to strict two-phase locking (S2PL)~\cite{bernstein86cc}, which is known to produce serializable executions.
S2PL acquires read and write locks on all data items in the transaction's data sets and releases them only after commit. 
We note that this is the behavior of transactions in our algorithm if they are not read-only. Thus, any non-read-only transaction in our algorithm can serialize. We now only have to argue about read-only transactions. Note that this was already covered in the argument for the algorithm in Section~\ref{sec:tweak-ORT}; since the client reads items in its read set before the validation phase, and then checks that all sequence numbers remain unchanged during the validation phase, these data items cannot have been modified during the time between the client's last read in the execution phase and its first read in the validation phase. Thus, read-only transactions can serialize during that time.

\subsection{Sacrificing seamless fault tolerance}

We now present an algorithm that satisfies all properties outlined in this paper except seamless fault tolerance. Note that this property only makes sense for the replicated case; unreplicated distributed systems do not tolerate any failures. Also, note that the algorithm only provides DAP, as opposed to DDAP, since we proved that achieving DDAP with the other properties is impossible when the system is sharded (\Cref{thm:dist-impossibility}).

Furthermore, note that seamless fault tolerance is closely related to fast decision. In particular, an algorithm that is fast deciding but does not satisfy seamless fault tolerance may stop satisfying the fast decision bounds not only if there are transaction conflicts, but also if there are node failures.

We capitalize on this fact in our solution. To create a serializable transactional system that satisfies all desirable properties except for even just $1$-seamless fault tolerance, we employ the base algorithm almost as-is; the only change is that this time, the client waits for \emph{all} nodes to reply, rather than just $n-f$. This solution is serializable, since each node handles all transactions, so conflicting transactions will compete for the same lock. 

To still allow fault tolerance (though not seamless), we allow the client to time out on the nodes. If the client times out while waiting to hear replies from all nodes, then the algorithm defaults to the one presented in Section~\ref{sec:tweak-ORT}. That is, the client sends a signal to all nodes letting them know it is switching gears, and restarts executing its current transaction, but this time waiting for just $n-f$ node responses, and using a two-round validation phase.
When a process receives such a restart signal from the client, it releases all locks it held for the current transaction.
The correctness of this algorithm follows directly from the correctness of the algorithm in Section~\ref{sec:tweak-ORT}.


\subsection{Sacrificing \DDAP{}}

Guaranteeing the rest of the properties without \DDAP{} is simple; we use the base algorithm, but maintain a global lock per node, rather than a lock per data item. Transactions always lock all nodes, even if they do not have any data item on some of them. The only exception is read-only transactions, which do not acquire locks, but check that the relevant lock is free and that the data items on the node have the same sequence number as they did in the execution phase. Clearly, this solution maintains fast decision and seamless fault tolerance, but does not preserve \DDAP. Note that since read-only transactions do not acquire locks, the weak invisible reads property is preserved. Furthermore, since all writing transactions acquire the locks of all nodes, transactions with the same write set execute non-trivial primitives on the same set of objects: all of the locks, and the sequence numbers and values of each data item in their write set. Thus, the invisible reads property is preserved as well.

The serializability argument for this algorithm is very similar to that of the algorithm in Section~\ref{sec:tweak-IR}. Non-read-only transactions acquire the locks on their entire data set, and commit only after successfully doing so. Thus, they are serializable by S2PL~\cite{bernstein86cc}. Read-only transactions can serialize sometime between the last read in the execution phase and the first in the validation phase.




\end{document}